\documentclass[fleqn,twoside]{article}
\usepackage{espcrc2}
\usepackage{epsfig}

\usepackage{graphicx}
\usepackage[figuresright]{rotating}

\newcommand{\AmS}{{\protect\the\textfont2
  A\kern-.1667em\lower.5ex\hbox{M}\kern-.125emS}}
\newcommand{\Z}{{\sf Z \!\!\! Z}}

\newcommand{\1}{{\sf 1 \!\! 1}}

\newcommand{\p}{\partial}

\title{Deconfinement $\!$in$\!$ Yang-Mills: a conjecture for a general gauge Lie group $G$}

\author{M. Pepe\address[Bern]{Institute for Theoretical Physics,
          Bern University, Sidlerstrasse 5, CH-3012 Bern,Switzerland.}}

\begin{document}

\begin{abstract}
Svetitsky and Yaffe have argued that --- if the deconfinement phase
transition of a $(d+1)$-dimensional Yang-Mills theory with gauge group $G$ is second
order --- it should be in the universality class of a $d$-dimensional scalar model
symmetric under the center $C(G)$ of $G$. These arguments have been investigated
numerically only considering Yang-Mills theory with gauge symmetry in the $G=SU(N)$ branch,
where $C(G)=\Z (N)$. The symplectic groups $Sp(N)$ provide another extension of 
$SU(2) = Sp(1)$ to general $N$ and they all have the same center $\Z(2)$. Hence, in contrast
to the $SU(N)$ case, $Sp(N)$ Yang-Mills theory allows to study the relevance of
the group size on the order of the deconfinement phase transition keeping the available
universality class fixed. Using lattice simulations, we present numerical 
results for the deconfinement phase transition in $Sp(2)$ and $Sp(3)$ Yang-Mills theories
both in $(2+1)$d and $(3+1)$d. We then make a conjecture on the order of the deconfinement
phase transition in Yang-Mills theories with general Lie groups $SU(N), SO(N), Sp(N)$ and
with exceptional groups $G(2), F(4), E(6),E(7),E(8)$. Numerical results for $G(2)$
Yang-Mills theory at finite temperature in $(3+1)$d are also presented. 
\vspace{1pc}
\end{abstract}

\maketitle

\section{Introduction and Overview}

The Yang-Mills theory with gauge group $G$ has a finite-temperature phase transition where
the confined colorless glueballs deconfine in a gluon plasma. This transition is signalled
by the spontaneous breaking of a global symmetry related to the center $C(G)$ of $G$. The
corresponding order parameter is the Polyakov loop  which transforms non-trivially 
$\Phi(\vec{x})' = z \Phi(\vec{x})$ under a global center transformation characterized by
the center element $z \in C(G)$. The expectation value of the Polyakov loop
$\langle \Phi \rangle = \exp(- \beta F)$ is related to the free energy $F$ of a 
static quark as a function of the inverse temperature $\beta = 1/T$. In the 
low-temperature confined phase the center symmetry is unbroken, i.e.\ 
$\langle \Phi \rangle = 0$, and hence the free energy of a single static quark 
is infinite. In the high-temperature deconfined phase, on the other hand, the 
center symmetry is spontaneously broken, i.e.\ $\langle \Phi \rangle \neq 0$, 
and the free energy of a quark is finite. Integrating out the spatial degrees of freedom
of the $(d+1)$-dimensional Yang-Mills theory, one can write down an effective action for
$\Phi$. It describes a scalar model with global symmetry $C(G)$ in $d$ dimensions. The
$C(G)$-symmetric confined phase of the gauge theory corresponds to the disordered phase of
the scalar model, while the $C(G)$-broken deconfined phase has its counterpart in the
ordered phase. Svetitsky and Yaffe~\cite{Sve82} argued that the interactions in the
effective description are short ranged. Hence, if the deconfinement phase transition is
second order, approaching criticality, the details of the underlying short-range
dynamics become irrelevant and only the center symmetry $C(G)$ and the dimensionality $d$ 
of space determine the universality class. Thus, one can exploit the universality of the
critical behavior to use a simple scalar model to obtain information about the much more
complicated Yang-Mills theory. On the other hand, if the phase transition is first order, the
correlation length does not diverge and there are no universal features. In this case the
$G$-symmetric Yang-Mills theory in $(d+1)$ dimensions and the $C(G)$-symmetric
$d$-dimensional scalar model do not share the same critical behaviour. 

For example, the $(d+1)$-dimensional $SU(2)$ Yang-Mills theory has center symmetry 
$\Z(2)$ and hence the effective theory is a $d$-dimensional $\Z(2)$-symmetric 
scalar field theory for the real-valued Polyakov loop. The simplest theory in 
this class is a $\Phi^4$ theory with the Euclidean action
\begin{equation}
S[\Phi] = \int d^dx \ \left[\frac{1}{2} \p_i \Phi \p_i \Phi + V(\Phi)\right].
\end{equation}
The scalar potential is given by $V(\Phi) = a \Phi^2 + b \Phi^4$,
where $b > 0$ for stability reasons. Indeed, for $a = 0$ this theory has a 
second order phase transition in the universality class of the $d$-dimensional 
Ising model. However, this does not guarantee that the deconfinement phase 
transition in $SU(2)$ Yang-Mills theory is second order. In particular, 
one could imagine an effective potential $V(\Phi) = a \Phi^2 + b \Phi^4 + c \Phi^6$,
involving also a $\Phi^6$ term which is marginally relevant in three dimensions.
The coefficient $c$ has to be positive in order to ensure that the 
potential is bounded from below, but the coefficient $b$ of the $\Phi^4$ term 
can now become negative. Then the phase transition becomes first order. 
The above argument extends straightforwardly to a generic gauge symmetry
group $G$ with center $C(G)$. Hence the order of the deconfinement transition is a
dynamical issue that can be addressed only by lattice simulations. 

The Yang-Mills theory on
the lattice is naturally formulated in terms of group elements while in the continuum the
fundamental field is the gauge potential, living in the algebra. An algebra can generate
different groups, however it is natural to expect that lattice Yang-Mills theories whose
gauge groups correspond to the same algebra have the same continuum limit.
Hence, instead of $SO(N)$ we consider its covering group $Spin(N)$.
Keeping this in mind, we look at the center subgroups $C(G)$ of
the various simple Lie groups $G$ 
\vskip-.6cm
\begin{eqnarray}
&&\hskip-.7cm C(SU(N))=\Z(N);\hskip.8cm C(Sp(N))=\Z(2) \\  
&&\hskip-.7cm C(SO(N))\hskip-.1cm\rightarrow \hskip-.1cm 
C(Spin(N))= \hskip-.1cm
\left\{\begin{array}{l} 
\hskip-.2cm\Z(2); \;\;\, N \;\mbox{odd} \\ 
\hskip-.2cm\Z(2)^2 ;\; N=4k \\
\hskip-.2cm\Z(4) ; \;\;\, N\hskip-.1cm=\hskip-.1cm4k\hskip-.1cm+\hskip-.1cm2 
\end{array}\right.\\
&&\hskip-.7cm C(G(2))= C(F(4))=C(E(8))=\{\1\}\\
&&\hskip-.7cm C(E(6))=\Z(3);\hskip.9cm C(E(7))=\Z(2)
\end{eqnarray}
\vskip-.2cm
Numerical simulations in $(2+1)$ and $(3+1)$ dimensions have been performed
for $SU(N)$ Yang-Mills theory in order to investigate the order of the
deconfinement transition and -- in case it is second order -- to
check the validity of the conjecture of Svetitsky and Yaffe. 
The currently known results are:\\ 
$\bullet$ {\underline {$(3+1)$ dimensions}}. 
The $SU(2)$ Yang-Mills theory has a second order deconfinement 
transition. Consistent with the Svetitsky-Yaffe conjecture, it is in the
universality class of the $3$d Ising model. $SU(N)$ Yang-Mills theories with 
$N=3,4,6,8$ have a first order deconfinement phase transition: hence there are no
universal features.\\ 
$\bullet$ {\underline {$(2+1)$ dimensions}}. Lowering the dimensionality of space makes
the fluctuations stronger. Indeed $SU(N)$ Yang-Mills theories with $N=2,3$ and, perhaps,
also $N=4$~\cite{deF03} have a second
order deconfinement phase transition. Also in these cases the Svetitsky-Yaffe conjecture
is verified and one finds that the $2$d universality classes are,
respectively, those of the Ising, 3-state Potts, and Ashkin-Teller models.

However the branch of $SU(N)$ groups is not a good choice to study the relation
between the order of the deconfinement phase transition and the size
of the group. In fact, in this case, when increasing the size 
$(N^2 - 1)$ of the group also the center $\Z(N)$ changes. 
In order to disentangle these two features we have considered the Yang-Mills theory with
gauge group $Sp(N)$. Since all $Sp(N)$ groups have the same center $\Z(2)$, we now
keep fixed the available universality class and we can directly study the
relevance of the size of the group on the order of the deconfinement transition. 
Finally, this also allows us to explore the deconfinement transition in Yang-Mills
theories with a gauge symmetry different from $SU(N)$. The results presented in these proceedings
are published in~\cite{Hol04}

\section{The Symplectic Group $Sp(N)$}

The group $Sp(N)$ is the subgroup of $SU(2N)$ which leaves the skew-symmetric 
matrix
\begin{equation}
J = \left(\begin{array}{cc} 0 & \1 \\ - \1 & 0 \end{array}\right) = 
i \sigma_2 \otimes \1,
\end{equation}
invariant. Here $\sigma_2$ is the imaginary Pauli matrix and $\1$ is the $N\!  \times\! N$
unit-matrix. The elements $U \in  SU(2N)$ belonging to  
$Sp(N)$ satisfy the constraint
\begin{equation}
\label{pseudoreal}
U^* = J U J^\dagger.
\end{equation}
Hence $U$ and $U^*$ are related by the unitary transformation $J$: consequently, the
$2N$-dimensional fundamental representation of $Sp(N)$ is pseudo-real. The matrix $J$
itself also belongs to $Sp(N)$, implying that in $Sp(N)$ Yang-Mills theory charge
conjugation is just a global gauge transformation. This property is familiar from 
$SU(2) = Sp(1)$ Yang-Mills  theory.

The constraint eq.(\ref{pseudoreal}) implies the following form of a generic
$Sp(N)$ matrix
\begin{equation}
\label{group}
U =  \left(\begin{array}{cc} W & X \\ - X^* & W^* \end{array}\right), 
\end{equation}
where $W$ and $X$ are complex $N \times N$ matrices such that 
$W W^\dagger + X X^\dagger = \1$ and $W X^T = X W^T$. Since center elements are multiples
of the unit-matrix, eq.(\ref{group}) implies $W = W^*\propto \1$: hence, the center of $Sp(N)$ is
$\Z(2)$. 

Writing $U = \exp(i H)$, where $H$ is a Hermitean traceless matrix,
eq.(\ref{pseudoreal}) implies that the generators $H$ of $Sp(N)$ satisfy the
constraint
\begin{equation}
\label{constraint}
H^* = - J H J^\dagger = J H J.
\end{equation}
This relation leads to the following generic form,
\begin{equation}
H =  \left(\begin{array}{cc} A & B \\ B^* & - A^* \end{array}\right),
\end{equation}
where $A$ and $B$ are $N \times N$ matrices such that $A = A^\dagger$ and $B = B^T$. 
Since $A$ and $B$ have, respectively, $N^2$ and $(N + 1)N$ degrees of
freedom, the dimension of the group $Sp(N)$ is $(2 N + 1)N$. 
There are $N$ generators of $Sp(N)$ that can be simultaneously diagonalized and so the rank
is $N$. The $N = 1$ case is equivalent to $SU(2)$, while the $N = 2$ case is equivalent to 
$SO(5)$, or more precisely to its universal covering group $Spin(5)$. 

\section{$Sp(N)$ Yang-Mills Theory on the Lattice}

The construction of $Sp(N)$ Yang-Mills theory on the lattice is 
straightforward. The links $U_{x,\mu} \in Sp(N)$ are group elements in the fundamental
$\{2N\}$ representation. We consider the  standard Wilson plaquette action
\begin{equation}
S[U] = - \frac{2}{g^2} \sum_P \mbox{Tr} \ U_P
\end{equation}
where $g$ is the bare gauge coupling and 
$U_P = U_{x,\mu} U_{x+\hat\mu,\nu} U^\dagger_{x+\hat\nu,\mu} U^\dagger_{x,\nu}$
is the plaquette. The partition function then takes the form
\begin{equation}
Z = \int {\cal D}U \exp(- S[U]),
\end{equation}
where the path integral measure is the product of the $Sp(N)$ Haar measures of each
link. Both the action and the measure are invariant under gauge transformations 
\begin{equation}
U'_{x,\mu} = \Omega_x U_{x,\mu} \Omega^\dagger_{x+\hat\mu},
\hskip 1cm \Omega_x \in Sp(N)
\end{equation}
The Polyakov loop is defined by
\begin{equation}
\Phi_{\vec x} = \mbox{Tr}({\cal P} \prod_{t = 1}^{N_t} U_{\vec x,t,d+1})
\end{equation}
where $N_t = 1/T$ is the extent of the lattice in Euclidean time, 
which determines the temperature $T$ in lattice units. 
The lattice action is invariant under global $\Z(2)$ center symmetry 
transformations $U_{\vec x,N_t,d+1}' = - U_{\vec x,N_t,d+1}$,
while the Polyakov loop changes sign, i.e.\ $\Phi_{\vec x}' = - \Phi_{\vec x}$.
As a consequence of the center symmetry, the expectation value of the Polyakov loop
always vanishes, i.e.\ $\langle \Phi \rangle = 0$, even in the
deconfined phase. This is simply because spontaneous symmetry breaking
does not occur in a finite volume. 
However, the Polyakov loop probability distribution $p(\Phi)$ 
does indeed allow one to distinguish confined from deconfined phases. In
the confined phase $p(\Phi)$ has a single maximum at $\Phi = 0$, while in the
deconfined phase it has two degenerate maxima at $\Phi \neq 0$. If the
deconfinement phase transition is first order, the confined and the two 
deconfined phases coexist and one can simultaneously observe three maxima 
close to the phase transition. At a second order phase transition, on the other
hand, the high- and low-temperature phases become indistinguishable. The two 
maxima of the deconfined phase merge and smoothly turn into the single 
maximum of the confined phase. Three coexisting maxima then do not occur in a 
large volume. 

As physical quantities useful in the finite-size scaling analysis 
presented below, we also introduce the Polyakov loop susceptibility
\begin{equation}
\chi = \sum_{\vec x} \langle \Phi_{\vec 0} \Phi_{\vec x} \rangle
= L^d \langle \Phi^2 \rangle,
\end{equation}
as well as the Binder cumulant~\cite{Bin81}
\begin{equation}
g_R = \frac{\langle \Phi^4 \rangle}{\langle \Phi^2 \rangle^2} - 3.
\end{equation}
The susceptibility measures the strength of fluctuations in the order parameter
while the Binder cumulant measures the deviation from a Gaussian distribution
of those fluctuations. We also consider the specific heat which takes the form
\begin{equation}\label{spec-heat}
C_V = 
\frac{1}{L^d N_t} (\langle S^2 \rangle - \langle S \rangle^2).
\end{equation}
In a finite volume $C_V$ has a maximum close to the critical coupling
$4N/g_c^2$ of the infinite 
volume theory. We denote the value of the specific heat at the maximum
by $C_V^{max}$. Another interesting observable is the latent heat 
\begin{equation}\label{lat-heat}
L_H = \frac{1}{L^d N_t}(\langle S \rangle_c - \langle S \rangle_d),
\end{equation}
which measures the difference of the expectation values of the action
in the confined and the deconfined phase. 
In the large volume limit $L_H$ and $C_V^{max}$ are related by~\cite{Bil92}  
\begin{equation}\label{speclat-heat}
C_V^{max} = L ^d N_t \frac{L_H^2 }{4}.
\end{equation}

\section{Numerical Results}
The $Sp(N)$ lattice Yang-Mills theory with the standard Wilson action can be
simulated using the Cabibbo-Marinari method~\cite{Cab82} of alternating
heat-bath~\cite{Cre80} and microcanonical overrelaxation~\cite{Adl81,Cre87,Bro87}
algorithms in various $SU(2) = Sp(1)$ subgroups.
In the next two subsections we report on the results of numerical simulations 
in $Sp(2)$ and $Sp(3)$ Yang-Mills theories in $(2+1)$d and $(3+1)$d. We denote with $L$
and $N_t$ the spatial and the temporal lattice extensions, respectively.
\subsection{$Sp(2)$ Yang-Mills Theory}
\begin{figure}[htb]
\begin{center}
\epsfig{file=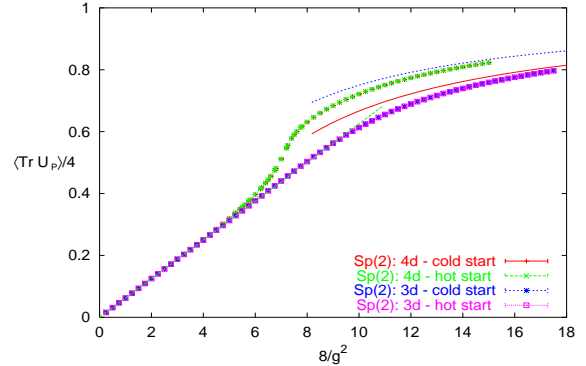,width=7.5cm,height=5.cm,angle=0}
\end{center}
\vskip-1.4cm
\caption{\it Monte Carlo data from hot and cold starts for the plaquette in 
$(2+1)$d and $(3+1)$d $Sp(2)$ Yang-Mills theory compared to analytic results 
in the weak and strong coupling limits.}\label{bulkSp2}
\vskip-.8cm
\end{figure}

As a first step we have scanned the expectation value of the plaquette in order to check
if the strong and the weak coupling regimes are separated by a bulk transition. Fortunately, 
we find no bulk phase transition that might interfere with the study of the deconfinement
transition both in $(2+1)$d and $(3+1)$d. Our Monte Carlo data for the expectation value of 
the plaquette are compared with analytic weak and strong coupling expansions in
figure~\ref{bulkSp2}.

\begin{figure}[h]
\vskip-.8cm
\begin{center}
\epsfig{file=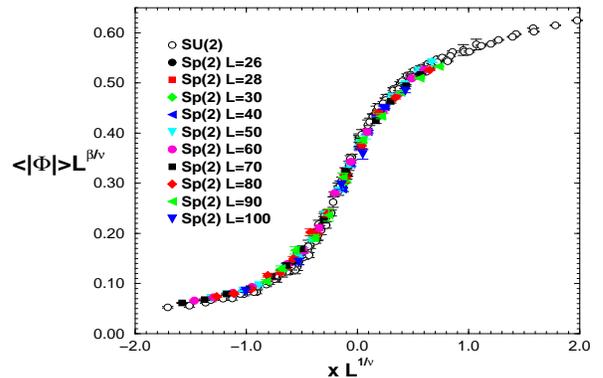,width=7.6cm,height=5.cm,angle=0}
\end{center}
\vskip-1.2cm
\caption{\it $(2+1)$d $Sp(2)$: finite-size scaling plot for 
$\langle |\Phi| \rangle  L^{\beta/\nu}$. Some $SU(2)$ data
are included too.}
\label{FSSchiSp2_2+1}
\vskip-.6cm
\end{figure}

\begin{figure}[htb]
\vskip-.8cm
\begin{center}
\epsfig{file=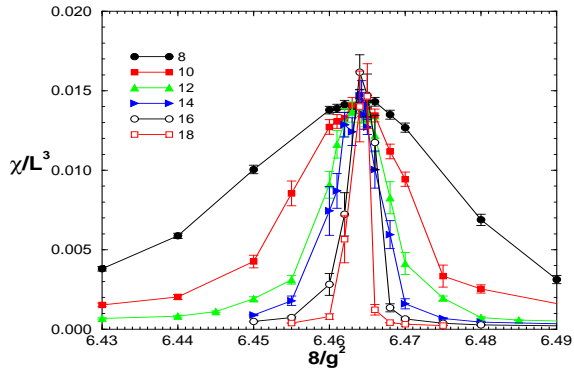,width=7.6cm,height=5.cm,angle=0}
\end{center}
\vskip-1.2cm
\caption{\it Scaling of $\chi$ in $Sp(2)$ Yang-Mills theory on $L^3\times 2$
  lattices . We estimate $8/g_c^2=6.4643(3)$.}
\label{FSSchiSp2_3+1}
\vskip-.8cm
\end{figure}

\begin{figure}[htb]
\begin{center}
\epsfig{file=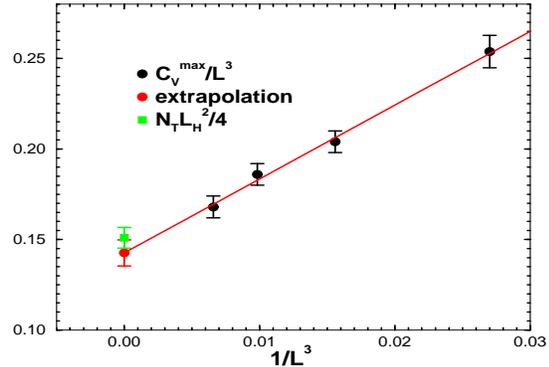,width=72mm,height=5.cm,angle=0}
\end{center}
\vskip-1.4cm
\caption{\it Approach to the infinite volume limit of $C_V^{max}/L^3$
for $(3+1)$d $Sp(2)$ Yang-Mills theory. The linear
extrapolation is in agreement with the measured latent heat $L_H$.}
\label{specheatSp2_3+1}
\vskip-.8cm
\end{figure}

In $(2+1)$d we observe a second order deconfinement transition, signalled by the
broadening of the probability distribution of $\Phi$ and, hence, by the increase of the
Polyakov loop susceptibility $\chi$ at criticality. A finite size scaling analysis
confirms the expectation that the universality class is that of the $2$d Ising model.
Figure~\ref{FSSchiSp2_2+1} shows the collapse on a single curve of the data for 
$\langle |\Phi | \rangle$ collected on lattices of different sizes $L^2\times 2$ and at
various couplings $g^2$. The variable $x=(g^2_c/g^2 -1)$ is a measure of the distance from
the critical coupling $g^2_c(N_t =2)$. The critical exponents $\nu$, $\beta$ and $\gamma$ are
characteristic of a universality class: in this case they have been fixed to those of the
2d Ising one. In figure~\ref{FSSchiSp2_2+1} we also plot rescaled data for
$SU(2)$ Yang-Mills theory in $(2+1)$d which has a second order deconfinement
transition in the 2d Ising universality class. The two sets agree excellently. 
In $(3+1)$d --~contrary to what one might have expected~-- the probability distribution of
$\Phi$ in the critical region clearly shows the coexistence of the symmetric and of the
broken phases. This indicates that the deconfinement transition is first order. 
Figure~\ref{FSSchiSp2_3+1} shows the susceptibility $\chi$ as a function of the gauge
coupling for different spatial sizes $L$, keeping $N_t=2$ fixed. It clearly turns out that, at
the critical coupling $8/g_c^2=6.4643(3)$, $\chi$ scales with the spatial volume $L^3$.
This quantitatively confirms the first order nature of the phase transition. 
Figure~\ref{specheatSp2_3+1} shows the maximum of the specific heat per volume,
$C_V^{max}/L^3$, as a function 
of the inverse volume $1/L^3$. The linear behavior is characteristic of a first order
phase transition. A linear extrapolation of $C_V^{max}/L^3$ to the infinite volume limit
(see eq.(\ref{speclat-heat})) is consistent with a direct measurement of 
$L_H$. This again supports the first order nature of the transition.

We have also checked if the deconfinement transition for $(3+1)$d $Sp(2)$ Yang-Mills
theory remains first order in the continuum limit. To test for scaling, we have measured
the dimensionless ratio $T_c/\sqrt{\sigma}$ between the deconfinement temperature $T_c$
and the square root of the string tension $\sigma$ at $T=0$. Our data indicate that we
are in the scaling regime, showing a scaling behaviour proportional to the lattice spacing
squared. 

\subsection{$Sp(3)$ Yang-Mills Theory}
\begin{figure}[htb]
\vskip-.1cm
\begin{center}
\epsfig{file=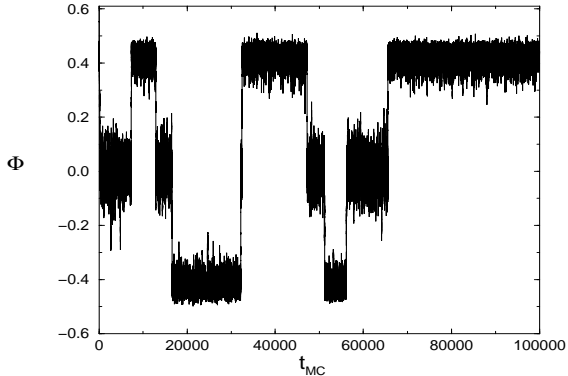,width=7.5cm,height=5.5cm,angle=0}
\end{center}
\vskip-1.6cm
\caption{\it $(3+1)$d $Sp(3)$: tunneling events between the symmetric and the broken
phases.}
\label{PolyhistorySp3_3+1}
\vskip-1.cm
\end{figure}

The results of $Sp(2)$ Yang-Mills theory show that in $(2+1)$d
fluctuations are stronger than in $(3+1)$d and the deconfinement
transition is second order. Expecting that the larger the group the
weaker the fluctuations, we have also investigated 
the deconfinement transition in $Sp(3)$ Yang-Mills theory. Consistent
with this picture, we find that in $(2+1)$ dimensions $Sp(3)$ Yang-Mills theory
has a first order deconfinement transition. The probability
distribution of $\Phi$ in the critical region indeed displays the
coexistence of the broken and of the symmetric phases.  
Equivalently, one can say that, since the number of colorless glueball states is almost
independent of the gauge group, the larger number of $Sp(3)$ gluons w.r.t. $Sp(2)$ gluons
in the deconfined phase, increases the difference between the relevant degrees of freedom
on the two sides of the phase transition. This drives deconfinement to take
place with a discontinuous first order transition. 
In $(3+1)$d -- similar to the $Sp(2)$ case -- $Sp(3)$ Yang-Mills
theory deconfines with a first order transition. 
Figure~\ref{PolyhistorySp3_3+1} shows tunneling events between the
coexisting symmetric and broken phases as a function of Monte
Carlo time $t_{\mbox{\tiny {MC}}}$. Finally, we note that $\!Sp(3)\!$ Yang-Mills theory 
has no bulk phase transition both in $(2\!+\!1)$d and $(3\!+\!1)$d.  
\vskip-.2cm

\section{Conjecture}
\begin{figure}[t]
\vskip.05cm
\begin{center}
\epsfig{file=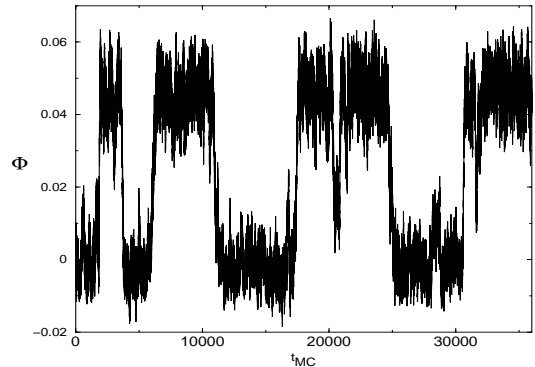,width=4.9cm,height=7.2cm,angle=-90}
\end{center}
\vskip-1.1cm
\caption{\it $(3+1)$d $G(2)$: Polyakov loop Monte Carlo history with tunneling events
  between two coexisting phases. The data refer to a simulation on a $18^3 \times
  6$ lattice. }
\label{G2fittemp}
\vskip-.9cm
\end{figure}

Our numerical results show that $(2+1)$d $Sp(2)$ Yang-Mills theory has a second
order deconfinement transition in the $2$d Ising universality
class. However, $(3+1)$d $Sp(2)$, $(2+1)$d and $(3+1)$d $Sp(3)$
Yang-Mills theories deconfine with a non-universal first order
transition. Hence, despite the fact that a universality class is available,
Yang-Mills theory can have a non-universal first order deconfinement
transition. A non-trivial center plays no role in determining the order of
this transition. Instead our $Sp(N)$ and the $SU(N)$ results
indicate that the order of the deconfinement transition is a
dynamical issue related to the size of the gauge group.
We conjecture that the difference in the number of the relevant
degrees of freedom between the confined phase (color singlet
glueballs) and the deconfined phase (gluon plasma) determines 
the order of the deconfinement transition. Thus, we expect that in $(3+1)$d
only $SU(2)$ Yang-Mills theory has a second order deconfinement
transition; in $(2+1)$d, due to stronger fluctuations, only $SU(N)$,
$N=2,3$ and, perhaps, $N=4$, and $Sp(2)$  Yang-Mills theories should have second order
transitions. According to this picture, $E(6)$ and $E(8)$ Yang-Mills
theories should also have a first order transition due to the large
size of the groups: 78 and 133 generators, respectively. 
For Yang-Mills theories with trivial center gauge groups $G(2)$, $F(4)$, $E(8)$~\cite{Hol03}
the large number of generators also suggests the presence of a first order transition even
if no symmetry can break down. Numerical results for $G(2)$ Yang-Mills theory in $(3+1)$d
indeed show the presence of a first order finite temperature phase transition. 
In figure~\ref{G2fittemp} we plot the Monte Carlo history of the Polyakov loop at finite 
temperature in the critical region: several tunneling events clearly show up. The probability
distribution has a two-peak structure indicating the presence of a first order finite
temperature transition.

\noindent{\it\bf {Acknowledgments}}.
It is a pleasure to thank the organizers of the ``QCD DOWN UNDER'' Workshop held in
Adelaide and in the Barossa Valley, for their kind
invitation. The results of the study presented in these proceedings have been obtained in
collaboration with K.~Holland and U.-J.~Wiese and they are published in~\cite{Hol04}.
For an extended and detailed list of references I also refer the reader to the published
version.  This work is supported by the Schweizerischer Nationalfonds.


\begin{thebibliography}{99}
\bibitem{Sve82}
B.~Svetitsky and L.~G.~Yaffe,
Nucl.\ Phys.\ B210 (1982) 423.

\bibitem{deF03}
P.~de Forcrand and O.~Jahn,
Nucl.\ Phys.\ Proc.\ Suppl.\  {\bf 129}, 709 (2004).

\bibitem{Hol04}
K.~Holland, M.~Pepe and U.~J.~Wiese,
hep- lat/0312022,to be published in Nucl.\ Phys.\ B.

\bibitem{Bin81}
K.~Binder, Z.\ Phys.\ B43 (1981) 119.

\bibitem{Bil92}
A.~Billoire, Int.\ J.\ Mod.\ Phys.\ C3 (1992) 913.

\bibitem{Cab82}
N.~Cabibbo and E.~Marinari,
Phys.\ Lett.\ B119 (1982) 387.

\bibitem{Cre80}
M.~Creutz,
Phys.\ Rev.\ D21 (1980) 2308.

\bibitem{Adl81}
S.~L.~Adler, Phys.\ Rev.\ D23 (1981) 2901; Phys.\ Rev.\ D37 (1988) 458.

\bibitem{Cre87}
M.~Creutz, Phys.\ Rev.\ D36 (1987) 515.

\bibitem{Bro87}
F.~R.~Brown and T.~J.~Woch, Phys.\ Rev.\ Lett.\ 58 (1987) 2394.

\bibitem{Hol03}
K.~Holland, P.~Minkowski, M.~Pepe, and U.-J.~Wiese, 
Nucl.\ Phys.\ B668 (2003) 207.
\end{thebibliography}
\end{document}